# Renormalization Group Functions for Two-Dimensional Phase Transitions: To the Problem of Singular Contributions

A. A. Pogorelov and I. M. Suslov

*Kapitza Institute for Physical Problems, Russian Academy of Sciences, ul. Kosygina 2, Moscow, 119334 Russia*
*e-mail: suslov@kapitza.ras.ru*

**Abstract**—According to the available publications, the field theoretical renormalization group approach in the two-dimensional case gives the critical exponents that differ from the known exact values. This property is associated with the existence of nonanalytic contributions in the renormalization group functions. The situation is analysed in this work using a new algorithm for summing divergent series that makes it possible to determine the dependence of the results for the critical exponents on the expansion coefficients for the renormalization group functions. It has been shown that the exact values of all the exponents can be obtained with a reasonable form of the coefficient functions. These functions have small nonmonotonities or inflections, which are poorly reproduced in natural interpolations. It is not necessary to assume the existence of singular contributions in the renormalization group functions.

## 1. INTRODUCTION

The field theoretical renormalization group approach [1] is the most appropriate method for calculating the critical exponents. In this method, phase transitions are described by the effective $\varphi^4$ theory with the action

$$S\{\vec{\varphi}\} = \int d^d x \left\{ \frac{1}{2}(\nabla \vec{\varphi})^2 + \frac{1}{2}\tau\vec{\varphi}^2 + \frac{1}{4}g\vec{\varphi}^4 \right\}, \qquad (1)$$

where $\vec{\varphi}$ is the $n$-component field vector, $d$ is the space dimension, $\tau$ is the distance to the transition, and $g$ is the coupling constant. The Fourier transform $G(k, \tau)$ of the correlation function $G(r, \tau) = \langle \varphi(x)\varphi(x + r) \rangle$ satisfies the Callan–Symanzik equation

$$\left[ -\frac{\partial}{\partial \ln k} + \beta(g)\frac{\partial}{\partial g} - \nu^{-1}(g)\frac{\partial}{\partial \ln \tau} - 2 + \eta(g) \right] \qquad (2)$$
$$\times G(k, \tau) = 0.$$

Under certain conditions, this equation has the solutions

$$G(0, \tau) \propto \tau^{-\gamma}, \quad G(k, 0) \propto k^{-2+\eta}$$

corresponding to the general phenomenology of the phase transitions [2]. The procedure reduces to determining a root of the equation $\beta(g^*)$ that specifies the stationary point of the renormalization group; after that, the main critical exponents are given by the expressions

$$\eta = \eta(g^*), \quad \nu = \nu(g^*), \qquad (3)$$
$$\gamma = \gamma(g^*), \quad \omega = \beta'(g^*),$$

while the other critical exponents are expressed through them using the well-known relations [2]. The renormalization group functions $\beta(g)$, $\eta(g)$, and $\nu^{-1}(g)$, as well as the functions

$$\eta^{(2)}(g) = \nu^{-1}(g) + \eta(g) - 2, \qquad (4)$$
$$\gamma^{-1}(g) = 1 + \eta^{(2)}(g)/(2 - \eta(g)),$$

are determined by the power series in the coupling constant $g$. The first expansion coefficients in these series are calculated by means of the diagrammatic technique, and the known Lipatov asymptotics exists for high orders [3, 4]. Owing to the factorial divergence of the series, their summation requires the use of special methods [5–7].

In the three-dimensional case, such an approach allowed the determination of the critical exponents with an accuracy to the third significant digit [5, 6, 8], but its application to two-dimensional systems appeared to be less successful (see table). In the pioneering work by Baker et al. [5], the four-loop expansions for the renormalization group functions at $n = 1$ were obtained and then summed by the Padé–Borel method. Owing to the comparatively low accuracy of the summation, the reasonable results were obtained for the "large" exponents ($\nu$, $\gamma$), whereas the "small" exponents ($\eta$, $\alpha$) remained almost undetermined (see the second column in table).

**Table**

|  | Exact value | [5] | [6] | [9] | This work |
|---|---|---|---|---|---|
| $\gamma$ | 1.75 | 1.72 ± 0.20 | 1.79 ± 0.09 | 1.790 | 1.785 ± 0.040 |
| $\nu$ | 1.00 | 0.92 ± 0.30 | 0.97 ± 0.08 | 0.966 | 1.01 ± 0.07 |
| $\eta$ | 0.25 | 0.08 ± 0.20 | 0.13 ± 0.07 | 0.146 | 0.145 ± 0.014 |
| $\eta^{(2)}$ | –0.75 | –0.83 ± 0.20 | –0.85 ± 0.07 | –0.872 | –0.865 ± 0.050 |
| $\omega$ | 4/3 (?) | 0.7 ± 0.4 | 1.3 ± 0.2 | 1.31 ± 0.03 | 1.345 ± 0.075 |
| $g^*$ | 1.754 (?) | 1.8 ± 0.3 | 1.85 ± 0.10 | 1.837 ± 0.030 | 1.82 ± 0.04 |

Le Guillou and Zinn-Justin [6] used a more accurate summation method (based on Borel transformation and conformal mapping) and carefully analysed uncertainty of the results, which allowed them to significantly reduce the error. As a result, the difference of $\eta$ and $\eta^{(2)} = \eta^{(2)}(g^*)$ from the known exact values was revealed for the two-dimensional Ising model (the third column in the table). More recently, Orlov and Sokolov [9] calculated the five-loop contributions in the renormalization group functions and found that the central values of the critical exponents almost did not change compared to [6] (the fourth column in table). Uncertainty of the results were not analysed in detail, but the character of their convergence with increasing the perturbation theory order provides a certain conclusion that the differences of the calculated exponents from the respective exact values are significant and the situation will not improved with the inclusion of further terms of the series. These differences were attributed to the existence of nonperturbative contributions with a singularity at zero [such as $\exp(-c/g)$] in the renormalization group functions. The summation results were independently confirmed by Calabrese et al. [10], who hypothesized the possibility of existing contributions singular at the stationary point $g^*$ [10, 11]. The existence of nonanalytic contributions means that the field theoretical renormalization group approach cannot provide the calculation of the critical exponents with an arbitrarily high accuracy; i.e., the usual belief that the problem is solved in principle is called in question.

The aim of this work is to analyse the situation using a new algorithm proposed for summing divergent series in [7, 12]. The application of this algorithm begins with the interpolation of the coefficient function, which makes it possible to almost completely remove the dependence of the results on variation in the summation procedure. Thus, only their dependence on the way of interpolation remains, which is directly related to the incompleteness of the initial information. Correspondingly, the relation of the summation results with the assumed behavior of the coefficient functions can be constructively analysed. For technical reasons, such a possibility was absent in other algorithms.[1] In fact, those algorithms were formulated in such a manner that unknown expansion coefficients were not considered clearly [5, 6]. This gives rise to the strong dependence of the results on variation in the summation procedure, which was restricted using semi-empirical recipes; for this reason, a reliable estimation of the accuracy becomes impossible.

Following the tradition [5, 6, 8, 9], apart the series for $\beta(g)$ we sum four series for functions $\eta(g)$, $\eta^{(2)}(g)$, $\gamma^{-1}(g)$, and $\nu^{-1}(g)$, among which only two functions are independent [see Eqs. (4)]. The latter sircumstance allows us to verify the self-consistency of the procedure. The last column of the table presents the exponents obtained for a certain set of "natural" interpolations (see Sections 2, 3). It is seen that uncertainty of the results for $\nu$ and $\gamma$ covers the exact value. Since these exponents can be taken as independent, there is no fundamental problem of the agreement of the obtained results with the exact values. However, the results obtained for $\eta$ and $\eta^{(2)}$ deviate from the exact values[2] (in accordance with the results of other authors [5, 6, 9, 10]), and there exists the technical problem of the inconsistency of the natural interpolations for different (interdependable) functions. Analysis shows (see Section 4) that the probable origin of this problem is the oscillatory behavior of the first expansion coefficients for $\eta(g)$ that generates a small nonmonotonicity or inflection in the coefficient functions for $\eta^{(2)}(g)$ and $\gamma^{-1}(g)$. These peculiarities are poorly reproduced by simple interpolations. As shown in Section 4, the exact values for the exponents $\nu$, $\gamma$, $\eta$, and $\eta^{(2)}$ can be obtained with a reasonable form of the coefficient functions. For this reason, there is no need to assume the singularity of the renormalization group functions, which contradicts the general principles (see Section 5).

---

[1] An important point is the stability of the algorithm with respect to smooth errors (including the umbiguity of interpolation); this circumstance makes it possible to avoid the catastrophic increase of errors in the course of resummation of the series [7, 12].

[2] One can see, that these deviations have no deep meaning, using the relations between the exponents. For the central values in the table, $\gamma = 1.785$, $\nu = 1.01$, and $\eta = 0.145$, the relation $\gamma = \nu(2 - \eta)$ is violated by 0.09, determining the scale of the uncontrolled systematic error, which is sufficient for the agreement of the results from the table with the exact values.

## 2. INITIAL INFORMATION AND SUMMATION PROCEDURE

The initial information is given by five-loop expansions for $\beta(g)$, $\eta(g)$, and $\eta^{(2)}(g)$, which are obtained in [9, 13], and series for $\gamma^{-1}(g)$ and $\nu^{-1}(g)$, which can be recalculated from them:[3]

$$\frac{\beta(g)}{2} = -g + g^2$$

$$-\frac{g^3}{(n+8)^2}(10.33501055n + 47.67505273)$$

$$+\frac{g^4}{(n+8)^3}(5.000275928n^2$$

$$+ 149.1518586n + 524.3766023)$$

$$-\frac{g^5}{(n+8)^4}(0.088842906n^3 + 179.6975910n^2 \quad (5)$$

$$+ 2611.154798n + 7591.108694)$$

$$+\frac{g^6}{(n+8)^5}(-0.00407946n^4 + 80.3096n^3$$

$$+ 5253.56n^2 + 53218.6n + 133972)$$

$$+ \ldots + c(-a)^N \Gamma(N+b) g^N + \ldots,$$

$$\eta^{(2)}(g) = -\frac{g}{(n+8)}(2n+4)$$

$$+\frac{g^2}{(n+8)^2}(n+2) \times 6.751257910 - \frac{g^3}{(n+8)^3}$$

$$\times (8.406683574n^2 + 65.16862270n + 96.71051110)$$

$$+\frac{g^4}{(n+8)^4}(0.583377094n^3 + 139.655555n^2 \quad (6)$$

$$+ 844.500099n + 1135.04499)$$

$$-\frac{g^5}{(n+8)^5}(-0.146720n^4 + 130.427n^3 + 2885.83n^2$$

$$+ 13691.4n + 16885.3)$$

$$+ \ldots + c'(-a)^N \Gamma(N+b') g^N + \ldots,$$

---

[3] Paper [9] contains an error in the five-loop term for $\gamma^{-1}(g)$, which was corrected in [13]. The five-loop contribution for $\eta^{(2)}(g)$ can be extracted from Table IV in [13] (where this function was denoted as $\eta t(g)$, whereas the first four loops (with a higher accuracy) can be obtained from the series for $\gamma^{-1}(g)$ and $\eta(g)$ presented in [9].

$$\eta(g) = \frac{g^2}{(n+8)^2}(n+2) \times 0.9170859698$$

$$-\frac{g^3}{(n+8)^2}(n+2) \times 0.05460897758$$

$$+\frac{g^4}{(n+8)^4}(-0.0926844583n^3 + 4.05641051n^2$$

$$+ 29.2511668n + 41.5352155) \quad (7)$$

$$-\frac{g^5}{(n+8)^5}(0.0709196n^4 + 1.05240n^3 + 57.7615n^2$$

$$+ 325.329n + 426.896)$$

$$+ \ldots + c''(-a)^N \Gamma(N+b'') g^N + \ldots,$$

$$\gamma^{-1}(g) = 1 - \frac{n+2}{n+8}g$$

$$+\frac{g^2}{(n+8)^2}(n+2) \times 3.375628955$$

$$-\frac{g^3}{(n+8)^3}(4.661884772n^2$$

$$+ 34.41848329n + 50.18942749)$$

$$+\frac{g^4}{(n+8)^4}(0.318993036n^3 + 71.70330240n^2 \quad (8)$$

$$+ 429.4244948n + 574.5877236)$$

$$-\frac{g^5}{(n+8)^5}(-0.119702n^4 + 69.3791n^3 + 1482.76n^2$$

$$+ 6953.61n + 8533.16)$$

$$+ \ldots + \frac{c'}{2}(-a)^N \Gamma(N+b') g^N + \ldots,$$

$$\nu^{-1}(g) = 2 - \frac{2(n+2)}{(n+8)}g$$

$$+\frac{g^2}{(n+8)^2}(5.834171940n + 11.66834388)$$

$$-\frac{g^3}{(n+8)^3}(8.352074597n^2$$

$$+ 64.62253293n + 95.83676746)$$

$$+\frac{g^4}{(n+8)^4}(0.676061553n^3 + 135.599145n^2 \quad (9)$$

$$+ 815.248932n + 1093.50978)$$

$$+ \frac{g^5}{(n+8)^5}(-0.217639n^4 + 129.375n^3 + 2828.07n^2$$

$$+ 13366.1 + 16458.4)$$

$$+ \ldots + c'(-a)^N \Gamma(N+b')g^N + \ldots,$$

as well as the parameters of the asymptotic behavior for high orders, which are calculated in [4]:

$$a = \frac{2.14793295333}{n+8}, \quad b = b' = \frac{n+7}{2},$$

$$b'' = \frac{n+5}{2},$$

$$c = 0.009838 \frac{n+8}{\Gamma(2+n/2)} \times 0.7335^n, \quad (10)$$

$$c' = 0.6441 \frac{n+2}{n+8} c, \quad c'' = 0.3306 c'.$$

The normalization of $g$ and $\beta(g)$ in Eqs. (5)–(9) is changed as compared to Eqs. (1) and (2) so that the first two coefficients in Eq. (5) are equal to unity (see [4, 5]).

The divergent series

$$W(g) = \sum_{N=N_0}^{\infty} W_N(-g)^N \quad (11)$$

whose coefficients $W_N$ have the asymptotic behavior $ca^N \Gamma(N+b)$ is summed by means of the Borel transformation

$$W(g) = \int_0^\infty dx\, e^{-x} x^{b_0-1} B(gx),$$

$$B(z) = \sum_{N=N_0}^{\infty} B_N(-z)^N, \quad B_N = \frac{W_N}{\Gamma(N+b_0)}, \quad (12)$$

where $b_0$ is an arbitrary parameter, and the subsequent conformal mapping (which is different from that in [6])

$$z = \frac{u}{(1-u)a}.$$

After that, reexpansion of $B(z)$ in the powers of $u$

$$B(z) = \sum_{N=0}^{\infty} B_N(-z)^N \bigg|_{z=f(u)} \longrightarrow B(u)$$

$$= \sum_{N=0}^{\infty} U_N u^N \quad (13)$$

gives the convergent series with the coefficients

$$U_0 = B_0, \quad U_N = \sum_{K=1}^{N} \frac{B_K}{a^K}(-1)^K C_{N-1}^{K-1}, \quad (14)$$

$$N \geq 1.$$

The asymptotic behavior of the coefficients $U_N$ for large $N$,

$$U_N = U_\infty N^{\alpha-1}, \quad N \longrightarrow \infty,$$

$$U_\infty = \frac{W_\infty}{a^\alpha \Gamma(\alpha) \Gamma(b_0+\alpha)}, \quad (15)$$

is related to the asymptotics for $W(g)$ in the strong coupling limit

$$W(g) = W_\infty g^\alpha, \quad g \longrightarrow \infty. \quad (16)$$

The coefficient function is interpolated by the formula

$$W_N = ca^N N^{\tilde{b}} \Gamma(N+b-\tilde{b}) \bigg\{ 1 + \frac{A_1}{N-\tilde{N}}$$

$$+ \frac{A_2}{(N-\tilde{N})^2} + \ldots + \frac{A_K}{(N-\tilde{N})^K} + \ldots \bigg\} \quad (17)$$

where the series is truncated and the coefficients $A_K$ are chosen from correspondence with the values of the coefficients $W_{L_0}, W_{L_0+1}, \ldots, W_L$, where $L_0$ needs not to coincide with $N_0$. The optimal value $\tilde{b} = b - 1/2$ [12] is used below, if another value is not indicated.

The parameter $\tilde{N}$ is used for varying the interpolation procedure. The coefficients $U_N$ for $N \leq N_{av} \approx 20$ are directly calculated by Eq. (14) and are then continued by power law (15) in order to avoid the catastrophic increase of errors [7, 12]. Thus, the consistent implementation of the algorithm necessarily requires the determination of the strong-coupling asymptotic behavior for $W(g)$ [see Eq. (16)]. For the summation of the series in the region $g \sim 1$, a high accuracy in the determination of this asymptotics is not necessary and its more or less detailed analysis implies the perspective applications to the strong-coupling region [7].

As compared to previous works [7, 12], a procedure for estimating errors in the region $g \sim 1$ is additionally developed. As a test example, we use the series for an anharmonic oscillator with the first nine coefficients. The best accuracy in determining the strong-coupling asymptotic behavior is reached at the optimal value $\tilde{N} = \tilde{N}_{opt} \approx 5.5$ [12]. The actual meaning of this value is clarified when considering the interpolation curves

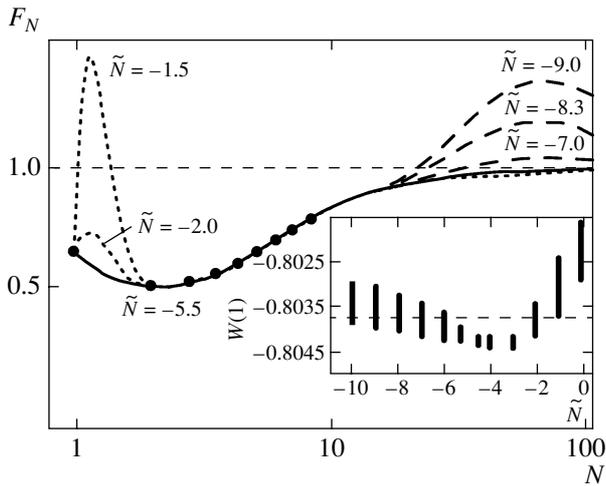

**Fig. 1.** Interpolation curves for the anharmonic oscillator at various $\tilde{N}$ values. The inset shows the results of summing the series for $g = 1$. Uncertainty of the results at a given $\tilde{N}$ value is associated with varying $b_0$ within the limits of uncertainty in the strong-coupling asymptotics. The horizontal dashed straight line in the inset is the exact value. The parameter $\tilde{N}_{av}$ is equal to 22.

(see Fig. 1) for the reduced coefficient function

$$F_N = \frac{W_N}{W_N^{as}} = \frac{W_N}{ca^N N^{b-1/2}\Gamma(N+1/2)} \qquad (18)$$

corresponding to the optimal parameterization of the Lipatov asymptotics with $\tilde{b} = b - 1/2$ [12]. When $\tilde{N}$ is varied near $\tilde{N}_{opt}$ by a value of the order of unity, the shape of the curves almost does not change. When $\tilde{N}$ is strongly decreased (for $\tilde{N} \lesssim -8$), nonmonotonicity appears in approaching $F_N$ to the limit value $F_\infty = 1$. When $\tilde{N}$ is strongly increased (for $\tilde{N} \gtrsim -2$), a large bump appears for noninteger $N$ values (see Fig. 1). These observations make it possible to separate the set of natural interpolations $-8.3 \leq \tilde{N} \leq -2.0$: the nonmonotonicity of the curves for them at large $N$ is at the level of the deviation of the last known point $F_L$ from 1 and the bumps at noninteger $N$ values are small as compared to a typical value $\bar{F}_N$. A particular choice of these constraints is not critical[4] and their reasonable variation weakly affects the result (see the inset in Fig. 1). The summation of the series for $g = 1$ on the set of natural interpolations provides the range $-W(1) = 0.8033 - 0.8042$, which contains the exact value $-W(1) = 0.80377$.

---

[4] The height of the maxima increases sharply with varying $\tilde{N}$, whereas the summation results are comparatively smooth functions of $\tilde{N}$; for this reason, the estimate of the uncertainty in the results is not too sensitive to the choice of the restrictions and is performed rather objectively.

## 3. SUMMATION RESULTS FOR NATURAL INTERPOLATIONS

### 3.1. Function $\beta(g)$

According to Eq. (5), the expansion coefficients $\beta_N$ with $N = 1, 2, …, 6$ are known. The interpolation by Eq. (17) with the use of all the coefficients ($L_0 = 1, L = 6$) provides the following conclusions. First, the interpolation curves for all $\tilde{N}$ are unsatisfactory in the sense of Section 2; i.e., they have essential bumps at noninteger $N$ or significant nonmonotonicities in the large $N$ region. Second, the attempt to estimate the strong-coupling asymptotics (16) gives unsatisfactory results: the pattern of the $\chi^2$ minima [12] is indistinct and poorly interpreted. We think that this occurs, because the value $F_N$ with $N = 1$ does not lie on a smooth curve obtained by analytic continuation from the points $N = 2, 3, 4, …$. Such a situation certainly takes place in the dimensional renormalization scheme, where $\beta(g) = -\epsilon g + \beta_0(g)$ for space of dimensionality $d = 4 - \epsilon$, and the function $\beta_0(g)$ refers to four-dimensional case (its expansion begins with $g^2$ and the coefficients are independent of $\epsilon$). Result (5) refers to another renormalization scheme, but a similar situation is also possible. For this reason, it is necessary to take $L_0 = 2$, i.e., to disregard the first point in the interpolation. More generally, possibility of such situations follows from the Sommerfeld–Watson summation procedure [12, Sect. 8.3]: the function $\mathcal{W}(z)$ that is the analytic continuation of $W_N$ onto the complex plane $[\mathcal{W}(N) = W_N, N = N_0, N_0 + 1, N_0 + 2, …]$ has a singular point $z = \alpha$, where $\alpha$ is the exponent in the strong-coupling asymptotic expression [see Eq. (16)]. If $\alpha$ is larger than $N_0$, one should take

$$W(g) = W_{N_0} g^{N_0} + … + W_{N_1} g^{N_1} + \tilde{W}(g), \qquad (19)$$

where $N_1$ is chosen from the condition $N_1 \leq \alpha < N_1 + 1$, sum the series for $\tilde{W}(g)$, and add the separated terms to the sum. Therefore, $N_1 = 1$ and $L_0 = 2$ should be taken for the function $\beta(g)$. After that, the results for the asymptotics (16) are satisfactory (see Fig. 2) and provide the value $\alpha \approx 1$, which a posteriori justifies the use of decomposition (19). It is important that the difference of $W_\infty$ from $-W_1$ lies beyond the error, so that the asymptotic behavior $W(g) \propto g$ is valid for the whole function $W(g)$.

Figure 3 shows the interpolation curves for $F_N$ obtained for various $\tilde{N}$ values. The inset shows the results for $g^*$ and $\omega$. It is easily seen that the natural interpolations correspond to the interval $-1.1 < \tilde{N} < 1.42$ and the summation of the series yields the result

$$g^* = 1.78 - 1.86 \qquad (20)$$

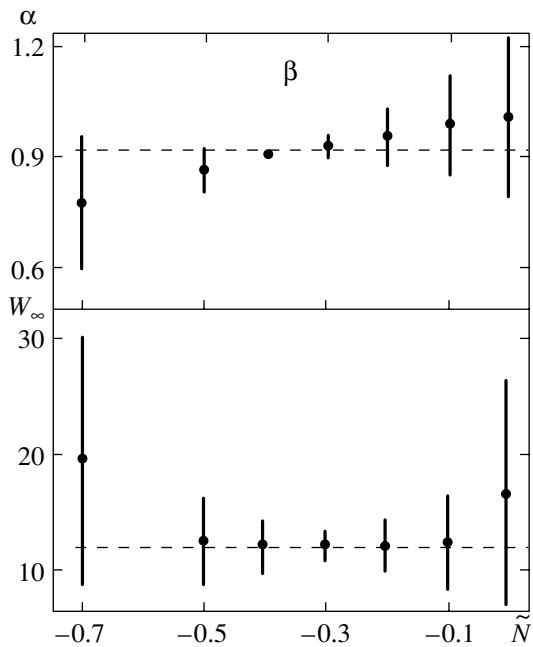

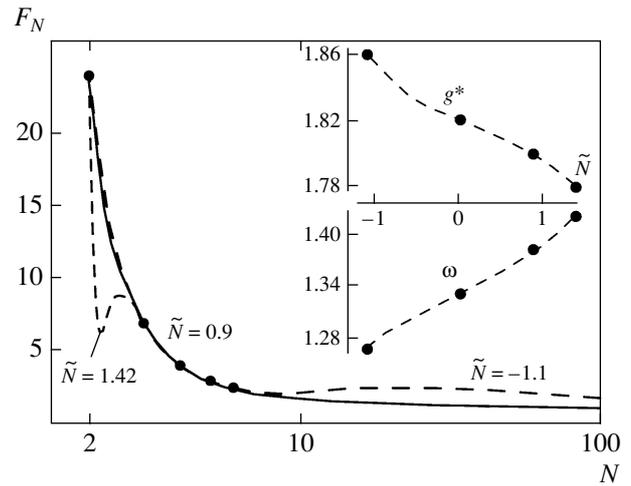

**Fig. 2.** Parameters of the strong-coupling asymptotics (16) for the function $\beta(g)$. Uncertainty of the results for a given $\tilde{N}$ value is determined according to [12]; $N_{av} = 21$.

**Fig. 3.** Interpolation curves for the function $\beta(g)$. The inset shows the results for $g^*$ and $\omega$.

for the root of the equation $\beta(g^*) = 0$. This result is usually compared to a value of 1.754 obtained from the analysis of the high-temperature series [14] (see table). Such a comparison is useful for orientation, but full coincidence of the results should not be expected. It is known that the renormalization group functions depend on the choice of the renormalization scheme [15] and only observables (critical exponents) are invariant. Result (20) is valid for field theoretical model (1) with the isotropic momentum cutoff $|k| < \Lambda$, whereas high-temperature series are constructed for lattice models, where the effective momentum cutoff is anisotropic. These details are physically insignificant, but they determine the difference between the renormalization schemes. As will be seen below, the results for the critical exponents do not favor any systematic shift of $g^*$, because a decrease in $g^*$ improves the results for $\gamma$ and $\eta^{(2)}$, but worsens the results for $\eta$. For this reason, the central value $g^* = 1.82$ seems to be the ideal compromise.

The numerical differentiation of $\beta(g)$ provides the results for the exponent $\omega$ presented in the table: they agree with the value $\omega = 4/3$, which appears in one of the versions of the conformal field theory, although other possibilities also exist (see discussion in [9]).

### 3.2. Function $\eta(g)$

According to Eq. (6), the expansion for $\eta(g)$ begins with $g^2$. Interpolation with $L_0 = 2$ (i.e., with the use of all the coefficients) implying the smoothness of $F_N$ for $N \geq 2$ provides unsatisfactory results (in terms of Section 2). For $L_0 = 3$, satisfactory interpolation curves are obtained for $1.2 < \tilde{N} < 2.9$ (see Fig. 4a); in this case, the parameters of the strong-coupling asymptotics are (see Fig. 5a)

$$\alpha = 2.01 \pm 0.01, \quad (21)$$
$$\tilde{W}_\infty = 0.051 \pm 0.007,$$

and indicate the existence of the singularity of $\mathcal{W}(z)$ at $z \approx 2$, which confirms the correctness of the removal of the first point. The results of series summation on the set of natural interpolations are shown in the inset in Fig. 4a. Their total uncertainty,

$$\eta = 0.131-0.158, \quad (22)$$

is small and does not include the exact value $\eta = 0.25$.

### 3.3. Functions $\eta^{(2)}(g)$, $\nu^{-1}(g)$, and $\gamma^{-1}(g)$

The interpolation with $L_0 = 1$ provides very uncertain results for the strong-coupling asymptotics (16). Therefore, it is reasonable to try to use the interpolation with $L_0 = 2$. In this case, according to Figs. 5b–5d,

$$\eta^{(2)}(g): \quad \alpha = 0.99 \pm 0.01, \quad \tilde{W}_\infty = 0.47 \pm 0.03, \quad (23)$$

$$\nu^{-1}(g): \quad \alpha = 0.985 \pm 0.115, \quad (24)$$
$$\tilde{W}_\infty = 0.345 \pm 0.235,$$

$$\gamma^{-1}(g): \quad \alpha = 0.955 \pm 0.105, \quad (25)$$
$$\tilde{W}_\infty = 0.22 \pm 0.11.$$

The natural interpolations correspond to the intervals $-2.5 < \tilde{N} < 1.8$, $-4.6 < \tilde{N} < 1.7$, and $-3.3 < \tilde{N} < 1.7$ for the functions $\eta^{(2)}(g)$, $\nu^{-1}(g)$, and $\gamma^{-1}(g)$, respectively; after that, the summation yields

$$\eta^{(2)} = -(0.835\text{--}0.879), \quad \nu^{-1}(g) = 0.980\text{--}1.060, \quad (26)$$
$$\gamma^{-1}(g) = 0.536\text{--}0.570.$$

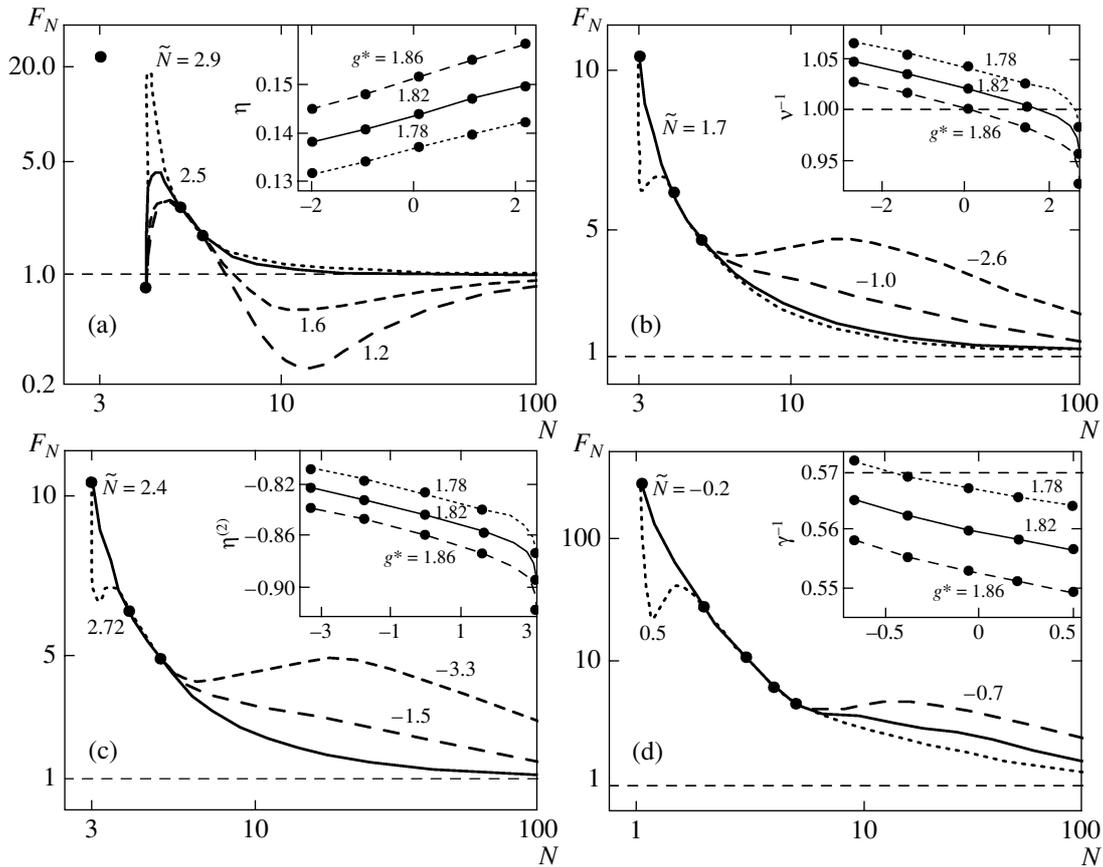

**Fig. 4.** Interpolation curves for the functions $\eta(g)$, $\nu^{-1}(g)$, $\eta^{(2)}(g)$ ($L_0 = 3$), and $\gamma^{-1}(g)$ ($L_0 = 1$). The inset shows the summation results at $g = g^*$.

According to Eq. (4), the asymptotic expressions for the functions under consideration do not contradict each other only when the function $\eta(g)$ is disregarded, because the expansion coefficients of the function $\eta(g)$ are much smaller than those for other functions; therefore, the difference between the reduced coefficient functions for $\eta^{(2)}(g)$, $\nu^{-1}(g)$, and $\gamma^{-1}(g)$ is very small and is not revealed by approximate analysis. The function $\eta(g)$ is very small for $g \sim 1$, but its asymptotics (16) grows more rapidly in the strong-coupling region [see Eq. (21)]. For this reason, the approximate linear behavior of $\eta^{(2)}(g)$, $\nu^{-1}(g)$, and $\gamma^{-1}(g)$ occurs at not too large $g$ values, whereas the function $\eta(g)$ distorts this behavior at large $g$ region.

According to Eq. (4), the general linear behavior of $\eta^{(2)}(g)$ and $\nu^{-1}(g)$ [see Eqs. (23) and (24)] is supplemented by additions proportional to $g^2$ with small coefficients, so that $\mathcal{W}(z)$ involves weak singularities at $z \approx$ 2, owing to which the interpolation with $L_0 = 2$ is invalid. At $L_0 = 3$, the interpolation curves for the functions $\eta^{(2)}(g)$ (see Fig. 4b) and $\nu^{-1}(g)$ (see Fig. 4c) have the satisfactory form for $-3.3 < \tilde{N} < 2.72$ and $-2.6 < \tilde{N} < 2.71$, respectively. The series summation results are shown in the inset in Figs. 4b, 4c and lie in the ranges

$$\eta^{(2)} = -(0.814\text{--}0.915), \quad \nu^{-1}(g) = 0.930\text{--}1.065. \quad (27)$$

The function $\gamma^{-1}(g)$ has an approximately linear behavior at $g \sim 1$, whereas its behavior at large $g$ region changes to approaching a constant or decrease [see Eq. (4)].[5] This indicates the absence of any singularities in the coefficient function for $N \geq 1$. Therefore, interpolation should be performed with $L_0 = 1$, which provides satisfactory results for $-0.7 < \tilde{N} < 0.5$ (see Fig. 4d); in this case, the summation of the series yields

$$\gamma^{-1} = 0.548\text{--}0.573. \quad (28)$$

---

[5] The decrease is possible when the asymptotic behavior of $\eta^{(2)}(g)$ is purely linear and does not include additions proportional to $g^2$; it looks rather probable in view of small uncertainty in Eq. (23).

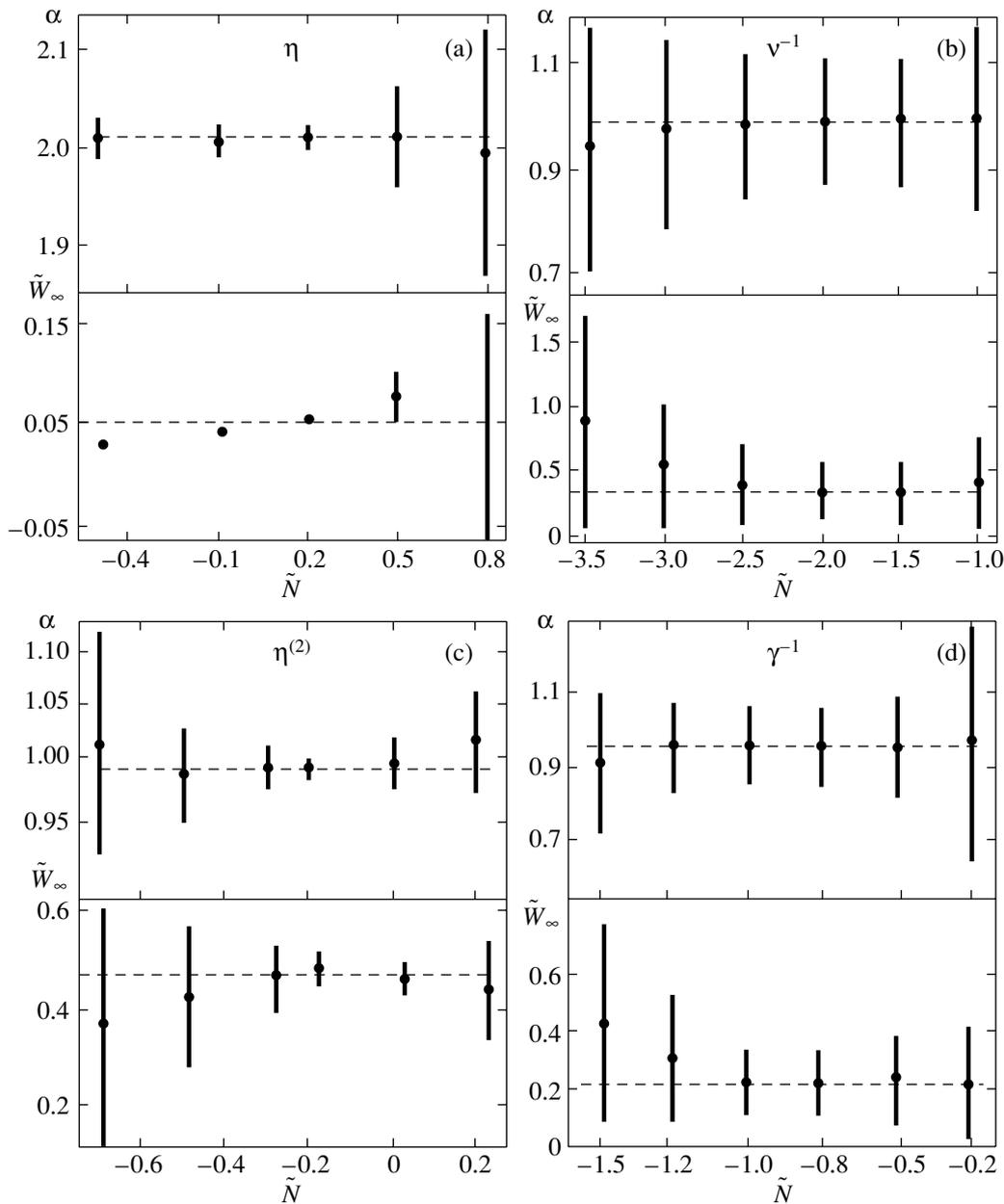

**Fig. 5.** Parameters of the strong-coupling asymptotic expression for the functions $\eta(g)$, $\nu^{-1}(g)$, $\eta^{(2)}(g)$, $\nu^{-1}(g)$, and $\gamma^{-1}(g)$ with $N_{av}$ = 20, 19, 18, and 19, respectively.

Comparison of Eq. (26) with Eqs. (27) and (28) shows that the coordination of the results for asymptotics shifts the central values for the exponents and leads to more adequate estimate of the accuracy. Uncertainty in $\nu^{-1}$ and $\gamma^{-1}$ on the set of natural interpolations [see Eqs. (27) and (28)] covers the exact values ($\nu^{-1} = 1$ and $\gamma^{-1} = 4/7 \approx 0.5714$); therefore, the fundamental problem of disagreement of the summation results with the exact values is absent. However, deviations from the exact values 0.25 and $-0.75$ for $\eta$ and $\eta^{(2)}$, respectively, are beyond the errors [see Eqs. (22) and (27)] and there exists a technical problem, revealing in not complete consistency of natural interpolations for different (interdependable) functions.

## 4. GUESSING OF THE COEFFICIENT FUNCTIONS

Let us verify if the exact values for all exponents could be ensured with a reasonable (and consistent) choice of the coefficient functions.

Let us begin with the function $\eta(g)$ whose first expansion coefficients have an oscillatory behavior (see Fig. 4a). For this behavior, the possibility of their rel-iable interpolation with high order asymptotic behavior seems to be doubtful from the very beginning. According to Section 3, the single-parameter set of interpolations used there cannot provide the exact value $\eta = \eta(g^*)$. Let us perform the simplest complication of the procedure and use the two-parameter family of interpolations. In this case, one parameter can ensure the exact $\eta$ value and the second parameter can be used to ensure the maximally "natural" form of the interpolation curve. Figure 6a shows the best results thus obtained. The solid line is obtained using Eq. (17) with two varying parameters $\tilde{N}$ and $\tilde{b}$ and corresponds to the values $\tilde{N} = -2.0$ and $\tilde{b} = -7.506$. The dashed line is obtained by varying $\tilde{N}$ and the first unknown coefficient $W_6$ (at $\tilde{b} = b - 1/2$) and corresponds to the values $\tilde{N} = 2.0$ and $W_6 = 0.0787$. In the latter case, clear and easily treated results corresponding to $\alpha \approx 2$ and $W_\infty \approx 0.7$ are obtained for the strong-coupling asymptotic behavior and confirm the correctness of the choice of $L_0$.[6] Both curves thus obtained (see Fig. 6a) are similar and characterized by oscillations with an amplitude of the same scale as the oscillations of the first coefficients: we think that such a behavior is admissible.

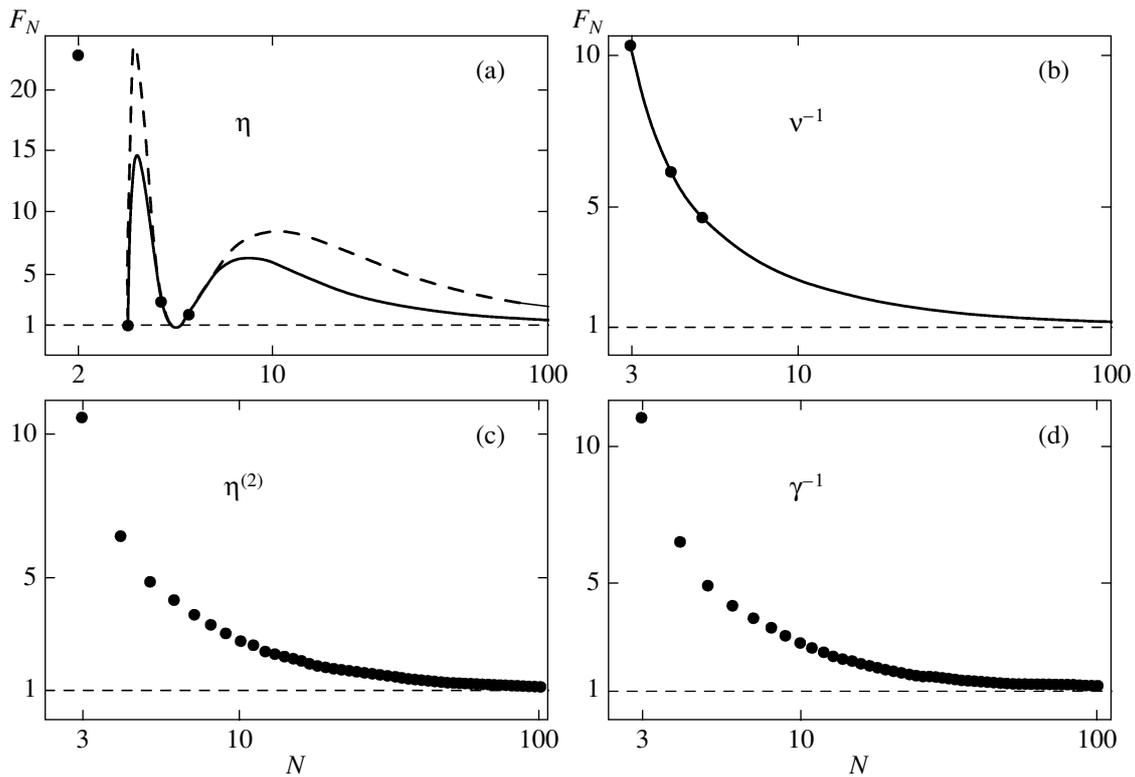

**Fig. 6.** Coefficient functions providing the exact values of the exponents.

---

[6] The interpolation with $L_0 = 3$ is used. The interpolation with $L_0 = 2$ lead to bumps in the interval $2 < N < 3$, indicating a pole at $N \geq 2$.

The function $\nu^{-1}(g)$ for which the exact value at $g^* = 1.82$ is reached for one of the natural interpolations (corresponding to $\tilde{N} = 1.77$) is used as the second independent function (see Fig. 6b). Accepting this interpolation curve for $\nu^{-1}(g)$ and the solid line in Fig. 6a for $\eta(g)$, one can use Eqs. (4) and obtain the expansion coefficients for $\eta^{(2)}(g)$ and $\gamma^{-1}(g)$, which are shown in Figs. 6c and 6d, respectively. The summation with these coefficients gives the exact values of $\eta^{(2)}$ and $\gamma^{-1}$. According to Eq. (4), the coefficient function for $\eta^{(2)}$ is the superposition of the smooth dependence for $\nu^{-1}(g)$ (see Fig. 6b) and the strongly oscillating dependence for $\eta(g)$ (see Fig. 6a), which gives rise to its inflection (see Fig. 6c) or nonmonotonicity (if more strong oscillations are allowed in Fig. 6a). Such an inflection is poorly described by simple interpolations; for this reason, $\eta^{(2)}$ obtained in Section 3, differs from the exact value. A similar inflection (Fig. 6d) or nonmonotonicity (Fig. 4d) is expected for the function $\gamma^{-1}(g)$.

Thus, the exact values of all four quantities $\eta$, $\nu$, $\eta^{(2)}$, and $\gamma$ can be reached with a quite reasonable choice of the coefficient functions. It is not necessary to assume the existence of singular contributions in the renormalization group functions [9–11].

## 5. THEORETICAL STATUS OF THE SINGULAR CONTRIBUTIONS

The assumption of the existence of nonperturbative contributions with a singularity at zero [such as $\exp(-c/g)$] [9] contradicts the general principles of the theory of divergent series. Indeed, the consistent treatment of divergent series is possible only in the framework of a certain set of axioms in which the Borel procedure for the series summation is accepted by definition [7]. Under this definition, no additions to the Borel integral such as $\exp(-c/g)$ are necessary, because such contributions have already been included in the definition of the sum. The problems appear only when the Borel integral is ill-defined [7]; however, this is not so in the present case, because all the singularities of the Borel images in the $\varphi^4$ theory are located on the negative semiaxis [16]. One can formally reject the possibility of the consistent treatment of divergent series and discard the indicated set of axioms, but such a position can hardly be expected from researchers involving in these investigations.

The possibility of contributions with a singularity at the fixed point $g^*$ [10, 11] is even more doubtful for the following reasons.

(i) The renormalization group functions in the Wilson picture [17, 18] relate the properties of finite blocks and should not have singularities, because the phase transitions are absent in the finite systems. In the general case, the regularity of the renormalization group functions is the basic principle of modern phenomenology replacing phenomenology of Landau theory: the latter explains the singular critical behavior on the basis of the analyticity of the thermodynamic potential, whereas the former phenomenology is based on the regularity of the renormalization group functions. The nonanalyticity of these functions means that the modern phenomenology is unsatisfactory, in contradiction with its obvious successes.

(ii) The constructive estimates indicate the power-law behavior of the Borel images at infinity (Section 3). In this case, Borel integral (12) determines a function regular on the positive semiaxis.

(iii) According to modern point of view, the critical exponents are continuous functions of the space dimension $d$, which in particular ensures the success of the Wilson $\epsilon$ expansion [17, 18]. In the dimensional regularization scheme, the Gell-Mann–Low function for $d = 4 - \epsilon$ has the form

$$\beta(g) = -\epsilon g + \beta_0(g),$$

where $\beta_0(g)$ is independent of $\epsilon$. The fixed point can be smoothly changed by varying $\epsilon$, whereas $\beta_0(g)$ cannot be singular at each point.

(iv) The constructive argumentation from [10, 11] is based on the comparison of corrections to scaling that follow from the general (many-parameter) version of the Wilson renormalization group with the analogous corrections in the field-theoretical formalism. However, such comparison is certainly incorrect, because the field-theoretical models are strictly renormalizable on the level of two parameters and do not involve corrections to scaling that are associated with the evolution of other parameters. The attempt to introduce them artificially gives rise to the necessity of assuming the singularity of the renormalization group function [10, 11].[7]

(v) The only real fact, to which the authors of [10, 11] appeal, is associated with the $1/n$ expansion. The first-order $1/n$ correction to the $\beta$-function has a form of the sum of the integrals over regular functions that can have singularities only in the case of their divergence. According to [10, 11], such divergences really exist due to the presence of small denominators and give rise to singularities of the type $(g - g^*)^\theta$. In fact, this result is associated only with the incorrectness of the $1/n$ expansion of integrands: the presence of small denominators shows that the real expansion parameter is not small in a certain part of the integration domain; for this reason, the restriction to the first order in $1/n$ is inappropriate.[8] A more accurate calculation inevitably results in the cutoff of the singularities found in [10, 11] and in the recovery of analyticity at $g = g^*$.

The above discussion indicates that the assumption of singular contributions made in [9–11] contradicts the general principles of the renormalization group approach and is theoretically unfounded. Recognition of their existence means that the field-theoretical renormalization group approach does not provide the principal solution to the problem of the critical exponents, because they cannot be calculated with arbitrarily high accuracy. On our opinion, this assumption is too strong and in fact there is no need in it: according to the above analysis, the exact values of the exponents can be obtained with a reasonable choice of the coefficient functions.

## ACKNOWLEDGMENTS

We are grateful to A.I. Sokolov for attracting our attention to this problem and for consultations on the questions related with the initial information. This work was supported by the Russian Foundation for Basic Research (project no. 06-02-17541).

---

[7] The correct stipulation: "Barring miracles, the approach [to the fixed point] should have nonzero components along any of the irrelevant directions" was present in [10, 11]. However, the field theoretical models certainly belong to miracles and have no projections onto the irrelevant directions, because they have a remarkable property of the strict renormalizability with two parameters.

[8] This possibly means that the renormalization group functions (in contrast to the critical exponents) in some renormalization schemes have no regular $1/n$ expansion.

*Translated by R. Tyapaev*